# Pressure-induced phase transitions and electronic properties $Cd_2V_2O_7$

*Daniel Díaz-Anichtchenko[1] and Daniel Errandonea[1,\*]*

[1]Departamento de Física Aplicada-ICMUV, Universidad de Valencia, Dr. Moliner 50, Burjassot, 46100 Valencia, Spain

*Corresponding author: daniel.errandonea@uv.es

## Abstract

We report a density-functional theory study of the structural and electronic properties of $Cd_2V_2O_7$ under high-pressure conditions. The calculations have been performed by using first-principle calculations with the CRYSTAL program. The occurrence of two structural phase transitions, at 0.3 and 10.9 GPa, is proposed. The crystal structure of the different high-pressure phases is reported. Interestingly a cubic pyrochlore-type structure is predicted to stabilize under compression. The two phase transitions involve substantial changes in the coordination polyhedra of Cd and V. We have also determined the compressibility and room-temperature equation of state of the three polymorphs of $Cd_2V_2O_7$. According to our systematic electronic band-structure calculations, at ambient conditions $Cd_2V_2O_7$ is an indirect wide band-gap material with a band-gap energy of 4.39 eV. In addition, the pressure dependence of the band gap has been determined. In particular, we have found that after the second phase transition the band gap decreases abruptly to a value of 2.56 eV.

## Introduction

Considerable attention has been paid to pyrovanadates during the last decade. Because of their band gap, appropriate photostability, luminescence, photocatalytic activity, ability to accept dopants, hydrophilicity, crystalline quality, thermal stability, and ability to absorb gases,[1] pyrovanadates are used in a wide range of applications.[1] Among them, cadmium pyrovanadate ($Cd_2V_2O_7$) has received remarkable attention because of its multiple possible technological uses. They include photocatalysis,[2-3] photoluminescence,[4] gas sensing,[5] among many others. For the optimization of these applications an accurate characterization of the physical properties is beneficial. In special, it is needed a good knowledge of the electronic and mechanical properties as well as of the structural stability under external stresses (e.g., pressure). Efforts for the precise characterization of electronic properties are also timely since there is a controversy in the literature regarding the band-gap energy ($E_g$), one of the key features for photocatalysis. In the literature, there are large discrepancies among the reported values for the band gap of $Cd_2V_2O_7$: 2.97,[2] 2.88,[3,6] 2.84,[7] and 2.47-1.77[4] eV. In addition, these values are all substantially smaller than the band-gap energy of $Mg_2V_2O_7$ ($E_g$ = 3.75 eV)[8] and $Zn_2V_2O_7$ ($E_g$ = 3.95 eV).[9] Band gap values smaller than 3 eV are unexpected in vanadates. In these compounds usually the band-gap energy is larger than 3.5 eV unless the divalent cation (Mg, Cd, Zn, Co, Cu, etc.) would have an electronic configuration with partially occupied d electrons,[10] which is not the case of Cd (electronic configuration $4d^{10}5s^2$). All the above-described facts indicate that further studies are needed to better know the value of $E_g$ and the nature of the electronic band gap. Band structure calculations at ambient and high-pressure could be helpful for it.[11]

Furthermore, the study of pyrovanadates under compression is a relevant topic for fundamental research. Materials science under high pressure (HP) is an important and very active research area opening new routes for stabilizing novel materials or original structures with tailored properties for specific applications. High pressure is also an essential tool for improving investigations on chemical bonding and consequently the physical and chemical properties of materials.[12] In contrast with other ternary oxides[13] little is known about pyrovanadates. Zinc pyrovanadate ($Zn_2V_2O_7$) undergoes a phase transition under very low pressure (0.7 GPa).[14] In addition, this compound undergoes two additional phase transitions at 3.0 and 10.2 GPa and it is more compressible than other vanadates[9,10,14] with a bulk modulus of 58 GPa. Secondly, in manganese pyrovanadate ($Mn_2V_2O_7$) high pressure influences its magnetic properties. In particular, an external pressure of 1.5 GPa induces a 100% enhancement of the Neel temperature.[15] On the other hand, the study of $Cd_2V_2O_7$ could be relevant for mineral physics because this vanadate is isostructural to the mineral thortveitite a silicate occurring in granitic pegmatites.[16] The crystal structure of $Cd_2V_2O_7$ is also shared by pyrogermanates, pyroarsenates, and pyrophosphates.[17] It consists of a pair of centrosymmetrically related corner-sharing $VO_4$ tetrahedral units linked to distorted $CdO_6$ octahedra. The understanding of the HP behavior of $Cd_2V_2O_7$ could give hints on the pressure behavior of all related oxides.

Density-functional theory (DFT) calculations have proven to be a quite efficient tool to study the HP behavior of ternary oxides.[9,18-21] In this work, we will use this method to characterize the different properties of $Cd_2V_2O_7$ and explore the possible existence of phase transitions. By considering the ambient pressure structure of $Cd_2V_2O_7$ described by space group (S.G.) C2/m[17], its metastable polymorph (S.G. $P\bar{1}$, γ-phase),[22] and other seven candidate structures, selected based on crystal-chemistry arguments, we have determined the possible occurrence of two subsequent phase transitions, and obtained the high-pressure structural sequence. In addition, to crystallographic information on the different structures, we will also report their compressibilities and equations of state. Furthermore, we will discuss the mechanisms driving the phase transitions. We will also report band structures and electronic densities of states, as well as the pressure dependence of the band-gap energy for the different polymorphs of $Cd_2V_2O_7$.

## Computational Details

Computer simulations have been performed using the periodic DFT framework using the CRYSTAL14 program package.[23] In the calculations the exchange correlation functional has been described using three different approximations; the Becke-Lee-Yang-Parr (B3LYP)[24-25] and Heyd-Scuseria-Ernzerhof (HSE06)[26] hybrid functionals as well as the widely used Perdew-Burke-Ernzerhof (PBE) functional.[27] For the calculations, Cd, V, and O atoms have been described by Cd_dou_1998, 86-411d3G, and 6-31d1G all electron basis sets, respectively. They have been taken from the Crystal website.[28] Using Pack–Monkhorst/Gilat shrinking factors: IS=ISP=4 we have controlled the diagonalization of the Fock matrix, that has been carried out at adequate fine *k*-point grids in the reciprocal space, which depend on the crystal structure (see below) of the phase under treatment. Thresholds controlling the accuracy of the calculation of Coulomb and exchange integrals have been set to $10^{-8}$ and $10^{-14}$ (and $10^{-16}$ for the pyrochlore structure, see below). This assures a convergence in total energy better than $10^{-7}$ hartree in all cases. The percent of Fock/Kohn-Sham matrices mixing has been set to 40 (IPMIX = 40). The choice of the exchange–correlation functional is of critical importance as it has a significant influence on the properties obtained.[29] Based upon our previous work in related $Zn_2V_2O_7$[9] and in the fact that B3LYP gives a more accurate description of the crystal structure of $Cd_2V_2O_7$ (see next section) at ambient pressure than HSE06 and PBE (see next section), calculations under high-pressure have been performed using the B3LYP functional.

In the calculations we have considered the known polymorphs of $Cd_2V_2O_7$ and structures isomorphic to other divalent-metal pyrovanadates, which are candidates for HP structures based on crystal chemistry arguments.[30] The structures included in the calculations are the ambient pressure phase $Cd_2V_2O_7$ (S. G. C2/m, β-phase)[17]; the metastable polymorph synthesized by the hydrothermal method at 493 K (S.G. P$\bar{1}$, γ-phase);[22] and the structures of α-$Zn_2V_2O_7$ (S.G. C2/c, α-phase),[31] $Hg_2V_2O_7$ (S.G. Pnma, δ-phase),[32] $Pb_2V_2O_7$ (S.G. P$2_1$/c, ε-phase),[33] $Mg_2V_2O_7$ (S.G. P$\bar{1}$, η-phase),[34] $Ni_2V_2O_7$ (S.G. P$2_1$/c, κ-phase),[35] and $Sr_2V_2O_7$ (S.G. P$\bar{1}$, ω-phase).[36] We have also contemplated a pyrochlore-type structure which has been previously reported in rare-earth pyrovanadates (S.G. Fd$\bar{3}$m, τ-phase).[37] From now on, we will use the names α, β, γ, δ, ε, η, κ, τ, and ω, respectively for denoting the different phases in the rest of the manuscript.

To consider long range Van der Waals interactions, we have included in calculations the semiempirical contributions proposed by Grimme,[38] which are especially useful to give an accurate description for metal oxides. The CRYSTAL program can perform an automated scan over the volume to compute energy (E) versus volume (V) curves that are then fitted to a third-order Birch-Murnaghan equation of state (EOS).[39] As a result, the pressure dependence of the crystal and electronic structure has been determined together with enthalpy vs. pressure curves for different candidate phases. These curves allow to determine the most stable phase as a function of pressure, the occurrence of phase transitions and transition pressures. The electronic density of states (DOS) and band structure have been calculated for different polymorphs based on the optimized geometries. All structures represented have been draw using VESTA.[40] Bond-distances and polyhedral parameters reported have been obtained with the same software.

## Results and discussion

**Crystal structure and phase transitions**

With the aim of testing different pseudopotentials, we have optimized the crystal structure of the stable polymorph of $Cd_2V_2O_7$ using B3LYP, HSE06, and PBE functionals. The obtained unit-cell parameters are shown in Table 1, where they are compared with reference parameters of the ambient pressure phase (β-$Cd_2V_2O_7$).[17] We conclude that B3LYP is the functional that best compare with experiments.[17] The error of volume calculated using the B3LYP, HSE06, and PBE functionals is respectively 2.7, 5.3, and 4%. The accuracy obtained with the B3LYP functional is typical of DFT calculations.[41] A similar underestimation of the calculated volume has been obtained for $Zn_2V_2O_7$ using the B3LYP functional.[9] A good agreement is also found between the atomic positions calculated with B3LYP and experiments.[17] This can be seen in Table 2. Since B3LYP describes reasonably well the ambient-pressure structure of $Cd_2V_2O_7$ and described properly the ambient- and high-pressure structures of $Zn_2V_2O_7$,[9] we will use the B3LYP functional for studying the high-pressure behavior of $Cd_2V_2O_7$.

The thermodynamically most-stable phase of $Cd_2V_2O_7$ at different pressures can be obtained by calculating the enthalpy versus pressures for different phases. In Fig. 1 we plot the enthalpy difference (ΔH) of different structure with respect to β-$Cd_2V_2O_7$. We note here that the calculated volume at zero pressure of β-phase is $V_0$=147.31 Å$^3$. if we use our calculated EoS (to be discussed in next subsection) and assume the experimentally determined volume at ambient pressure, this volume corresponds to a pressure of -4 GPa. Thus, as assumed in related compounds,[42] in the discussion of results we have considered there is an offset of 4 GPa between calculated ($P_{calc}$) and real pressures. In Fig. 1 and other figures, for the sake of transparency, we present both the calculated pressure ($P_{calc}$), in the bottom horizontal axis, and the corrected pressure (P), in the top horizontal axis. As pressure increase, ΔH between γ-$Cd_2V_2O_7$ and β-$Cd_2V_2O_7$ decreases. At a calculated "pressure" of -3.7 GPa, the γ-phase (a metastable polymorph at ambient conditions[22]) becomes the lowest enthalpy phase, which supports the occurrence of the β–γ transition at this pressure. With the 4 GPa offset mentioned above, the transition pressure is 0.3 GPa. This transition pressure is comparable to the pressure of the phase transition observed in $Zn_2V_2O_7$.[9,14] The calculated structure of γ-$Cd_2V_2O_7$ is compared with experiments[22] in Tables 1 and 2. The agreement is reasonable with only an underestimation of the calculated volume of 2%. This volume underestimation corresponds also to a pressure of 4 GPa according to the EoS calculated for the γ-phase (to be discussed in next subsection). Calculations also predict a second phase transition at $P_{calc}$ = 6.9 GPa from the γ-phase to the τ-phase, which with the assumed pressure offset leads to a transition pressure of 10.9 GPa. This cubic phase is the lowest enthalpy phase among the eight considered phases up to 14 GPa. The structural information of τ-phase can be found in Tables 1 and 2.

We will discuss now the proposed structural sequence. We will start describing β-Cd$_2$V$_2$O$_7$. Fig. 2(a) shows a perspective of the crystal structure of Cd$_2$V$_2$O$_7$ at ambient pressure (β-phase). The unit cell contains 2 formula units. It consists of layers of CdO$_6$ irregular octahedra. The layers are parallel to the (001) plane. This is shown in Figs. 2(b) and 2(c). The layers are formed by edge-sharing octahedra making a hexagon with an empty space in the center. This can be seen in Fig. 2(d). The layers are interconnected by isolated V$_2$O$_7$ dimers, each of them formed by two corner sharing VO$_4$ tetrahedra. Each VO$_7$ dimer, shown in Fig. 2(e), is a linear molecule with a V-O-V angle of 180º. The linear molecule points along the [10, 0, $\bar{3}$] direction, i.e., it is aligned parallel to the b-axis. According to the behavior of other vanadates, the large CdO$_6$ octahedron (octahedral volume 14.26 Å$^3$) is more compressible than the small and rigid VO$_4$ tetrahedron (tetrahedral volume 2.52 Å$^3$).[13] The difference of compressibility is due to the differences between Cd-O and V-O bond distances.[13] Compression of the b-axis involves only the compression of the CdO$_6$ layers while the compression along the other axes will also involve a compression of the VO$_7$ dimmers. Thus, the b-axis is expected to be the most compressible axis. This conclusion is supported by our calculations as we will show when discussing the pressure dependence of unit-cell parameters.

We will comment now on structural changes associated to the phase transition from β-Cd$_2$V$_2$O$_7$ to γ-Cd$_2$V$_2$O$_7$. As it can be seen it Table 3 this transition involves changes in the coordination polyhedra. The crystal structure γ-Cd$_2$V$_2$O$_7$ is represented in Figs. 3a, 3b, 3c, and 3d. The structure is isotropic with Ca$_2$V$_2$O$_7$.[43] In the structure there two independent Cd atoms, one is hexa-coordinated and the other hepta-coordinated. According to the distortion index,[40] these units are more regular than the CdO$_6$ octahedra of β-Cd$_2$V$_2$O$_7$. In γ-Cd$_2$V$_2$O$_7$, CdO$_6$ and CdO$_7$ polyhedra are edge connected making a three-dimensional framework structure that is the main building unit of the crystal structure. On the other hand, V atoms also occupy two independent positions. One is in tetrahedral coordination and the other in a distorted trigonal–pyramidal coordination environment. As shown in Fig, 3e, two VO$_5$ pyramids share edges and each of these units is connected to a VO$_4$ tetrahedron sharing a corner atom. This way, VO$_4$ and VO$_5$ units form an isolated V$_4$O$_{14}^{8-}$ anion. Clearly the phase transition involves the formation of new bond. As we will show in the next section, it also involves a volume collapse. Therefore, γ-Cd$_2$V$_2$O$_7$ is denser than β-Cd$_2$V$_2$O$_7$.

At the second phase transition the symmetry of Cd$_2$V$_2$O$_7$ is enhanced and the coordination number of Cd and V too (see Table 3). τ-Cd$_2$V$_2$O$_7$ has a pyrochlore crystal structure which is represented in Figs. 4a and 4b. In this structure the Cd cation is eight-fold coordinated making a near regular dodecahedron and the V atom is six-fold coordinated forming a platonic octahedron. The Cd and V are ordered along the [110] direction. This structure has been reported before only for rare-earth vanadates (e.g., Y$_2$V$_2$O$_7$)[37] where V is tetravalent, but never observed in pyrovanadates where V is pentavalent like Cd$_2$V$_2$O$_7$. Thus, the pressure-induced stabilization of τ-Cd$_2$V$_2$O$_7$ suggests a charge transfer from V to Cd atoms. It is known that at ambient conditions A$_2$X$_2$O$_7$ pyrochlores are stable only if the of ratio of radii of A cation (r$_A$) and X cation (r$_B$), r$_A$/r$_B$ is in the range from 1.46 to 1.78.[44] In Cd$_2$V$_2$O$_7$, r$_A$/r$_B$ = 2.70, which is far away from the expected value for a stable pyrochlore structure. However, if the charge transfer is assumed, r$_A$/r$_B$ is equal to 1.63; which is compatible with a pyrochlore structure. Thus, a pressure-induced charge transfer appears to be a possible hypothesis to explain the occurrence of the second phase transition.

In summary, the application to external pressure to $Cd_2V_2O_7$ favors an increase of cation coordination. In particular, the coordination number of Cd goes from 6 to 6/7 and then to 8 following the predicted phase transitions; and the coordination number of V goes from 4 to 4/5 and then to 6. In all the phases the coordination polyhedra are convex polyhedra, with a Euler characteristic of 2; i.e.; in the three phase coordination polyhedra are topologically equivalent to the sphere. The formation of pyrochlore-type τ-$Cd_2V_2O_7$ under high-pressure is reasonable according to the Voronoi polyhedral analysis[45] which predicts that under compression the number of phases of coordination polyhedra should increase. Volume discontinuities and coordination changes at the phase transitions support that they are first-order reconstructive phase transitions.

**Equations of state and compressibility**

Now we will discuss the axial volume compressibility of the different phases. The pressure dependence of unit-cell parameters and volume are represented in Figs. 5 and 6. The two transitions involve large volume discontinuities as can be seen in Fig. 6. This means that both transitions are first-order in nature. The results for pressure dependence of the volume have been fitted with a third order Birch–Murnaghan equation of state. The obtained bulk modulus ($B_0$), its pressure derivative ($B_0'$), and the zero-pressure volume ($V_0$) are given in Table 4. The obtained bulk modulus for β-$Cd_2V_2O_7$ (80.9 GPa) is between the values previously reported for $Zn_2V_2O_7$ (58 GPa)[14] and $Co_2V_2O_7$ (104 GPa).[46] As a first approximation, it can be expected that in our compound the volume reduction of the unit-cell corresponds to the compression of $CdO_6$ octahedra.[30] Under this hypothesis, using the empirical relationship proposed by Errandonea and Manjon[30] a bulk modulus of 85(10) GPa is estimated for the β-phase of $Cd_2V_2O_7$. This estimation is in good agreement with the results of our calculations. For the HP phases the bulk modulus increases as the unit-cell volume decreases, which agrees with the systematic followed by oxides.[47] The bulk modulus obtained for τ-$Cd_2V_2O_7$ (226 GPa) is similar to bulk moduli reported in other pyrochlore type oxides.[48,49]

From Fig. 5 it can be seen that the response to pressure of the β- and γ-phases is anisotropic. This can be clearly observed in Fig. 5 for the β- and γ-phases, where the evolution of parameters is quite different, because the different linear compressibilities. To know the magnitudes and directions of the principal axes of compression we have calculated the eigenvalues and eigenvectors of the compressibility tensor. We have obtained them for β- and γ-phases using PASCAL.[50] For the τ-phase the response is isotropic due to the high symmetry. In this case the linear compressibility has been calculated directly from the pressure dependence of the unit-cell parameter. The values of the main-axis compressibilities are provided in Table 5. We have found that in the β-phase, the major compression direction is parallel to the [010] crystallographic axis, as we expected seeing the Fig. 1, where this direction has much empty space between the layers of vanadium polyhedra. The minimum compressibility is in the orthogonal direction, approximately in the [5 0 11] direction, being the linear compressibility along this direction 3/10 times the one along the b-axis. There is a negative linear compressibility in the [101] direction, which means that the crystal expands in that direction. In the γ-phase, we have found that two of the main directions ([3,4,2] and [2,1,-3]) with similar compressibilities. In the third direction ([14 − 2 − 1]) the linear compressibility is extremely small, comparable to that of zero-linear compressibility materials.[51]

**Pressure effects on the band structure**

Now we will present the results of our band-structure and electronic density of state (DOS) calculations. The calculated band structure and electronic DOS (using the B3LYP hybrid functional) for the different phases are shown in Fig. 7. We have found that the three main phases have indirect band gap. The β-phase has a calculated band gap of 4.39 eV, with the top of the valence band (VB) at the (100) point of the Brillouin zone (BZ) and the bottom of the conduction band (CB) at (001). The γ-phase has a calculated band gap of 4.55 eV, with the top of the valence band at the $\Gamma$ point of the BZ and the bottom of the conduction band at (011). The τ-phase has a calculated band gap of 2.55 eV, with the top of the valence band at the (010) point of the BZ and the bottom of the conduction band at $\Gamma$ point. In the case of the β-phase, we have also obtained the values of the band gap with the HSE06 and PBE functional, 4.23 and 2.56 respectively. The comparison between the band-gap energies calculated with the different functionals follows a similar trend that in a previous work on $Zn_2V_2O_7$.[9] HSE06 which usually gives an accurate description of the band-gap energy[52] give a value of Eg that only differs by 3.5% from the value obtained using B3LYP (4.39 eV). In contrast, PBE gives a considerably smaller value. This fact is not surprizing since PBE usually underestimates Eg.[52,53]

Next, we will compare our results for β-$Cd_2V_2O_7$ with the literature. As we mentioned before, there are discrepancies between experimental results and the band gap calculated with the B3LYP functional, some experimental results are 2.97[2], 2.88,[3] 2.84,[7] and 2.47-1.77[4] eV. These values are all more than 30% smaller than the calculated energy using HSE06 and B3LYP functionals. In addition, they compare well with the PBE predictions, which usually underestimate Eg in vanadates.[9,54] We conclude than in previous studies the band-gap energy has been underestimated. We provide next arguments supporting our hypothesis. All previous experiments have been carried out in nanocrystalline or powder samples. In them, the band gap has been determined either from photo-acoustic spectroscopy[4] or from diffuse-reflectance measurements.[2,3,7] In addition, diffuse reflectance measurements have been performed rather than transmittance measurement in single crystals. Therefore, the band gaps have been only estimated in previous experiments as concluded in previous works.[55,56] These estimations have been based on the Kubelka–Munk method and Tauc plot analyses. Both techniques and the use of powder or nano-powder samples usually leads to an underestimation of Eg.[57] This is due to the fact, that Urbach tail absorptions and artifacts related to diffuse light mask the onset of the fundamental absorption associated to the band gap. Many of these issues have been discussed in detail in the method have been analyzed in a relatively a good review.[58] Interestingly, in two of the previous works, after the onset of the first absorption a second absorption, which is steeper than the first absorption is observed at 4.2 eV.[2] Quite probably, this second absorption corresponds to the fundamental band gap as recently shown for $ScNbO_4$.[59] A band gap of 4.2 eV is in full agreement with our HSE and B3LYP calculations. This value of the band-gap energy is also in agreement with DFT calculations performed using the generalized-gradient approximation.[60]

Now we will focus on the orbital composition of the band structure of β-$Cd_2V_2O_7$. In the DOS shown in Fig. 7, it can be seen that the bottom of the CB is dominated by V 3d orbitals, with a small contribution of O 2p states. On the other hand, the top of the VB is dominated by O 2p orbitals and the contribution from other atoms is negligible. This feature is common is a fingerprint of most vanadates.[9,19,61,62] In particular of $Zn_2V_2O_7$.[9] The fact the V 3d and O 2p orbitals dominate the states near the Fermi level explain why compounds like $Cd_2V_2O_7$ of $Zn_2V_2O_7$ have a wide band gap. In the HP phases, the band structure is modified, due to the modification of the crystal structure. However, the states near the Fermi level are dominated by V 3d and O 2p orbitals too. This can be seen in Fig. 7.

To conclude we will analyse the evolution of the energy gap with the pressure. We have found that at the first transition the band gap increases to 4.55 eV. This is a consequence of the enhancement of the splitting between bonding and anti-bonding states. At the second transition, the band gap suddenly decreases to 2.55 eV. This mainly due to the increase of hybridization between V 3d and O 2p electrons. In Fig. 8 we present the pressure dependence of the band-gap energy for the different phases. In the small pressure range of stability of the β-phase the band-gap energy remains nearly constant. In contrast, in both the γ- and τ- phases the band-gap energy slightly increases with pressure. Unfortunately, there are no experiments to compare with. We hope our results will trigger optical-absorption experiments on $Cd_2V_2O7$ using single crystals to accurately determine the pressure dependence of the band gap and compare with our predictions.

## Conclusions

In this work using of density-functional theory calculations we have studied the high-pressure behavior of cadmium pyrovanadate ($Cd_2V_2O_7$). We have found that the B3LYP is the functional that describe better the properties of $Cd_2V_2O_7$. Our calculations predict the existence of two phase transitions for pressures smaller than 14 GPa. The first a transition takes place at 0.3 GPa from β-$Cd_2V_2O_7$ to metastable γ-$Cd_2V_2O_7$. Then a transition occurs at 10.9 GPa to pyrochlore-type τ-$Cd_2V_2O_7$. Changes induced in the crystal structure have been analyzed and cation coordination changes have been found to be related to the phase transitions. It has been also found that pressure not only increases the density of $Cd_2V_2O_7$, but also makes it symmetric. The compressibility of the different phases has been also studied, being found that the compressibility of $Cd_2V_2O_7$ is between that of $Zn_2V_2O_7$ and $Co_2V_2O_7$. In addition, the response to pressure is found to be non-isotropic in the β- and γ-phase, but isotropic the τ-phase. The main axis of isothermal compressibility tensor has been determined for every phase. Regarding electronic properties, the band-structure and electronic density of states of the different phases have been obtained. We conclude that in contrast with previous studies, $Cd_2V_2O_7$ is a wide band-gap material (4.2-4.39 eV). Reasons for previous underestimations of the band-gap energy have been discussed. At the transition to the τ-phase there is a band-gap collapse, becoming the band-gap energy 2.56 eV. This is mainly caused by the enhance of hybridization between states of V 3d and O 2p electrons. Finally, the pressure dependence of the band-gap energy is reported for each phase. Explanations for the observed phenomena are provided with the comparison of related compounds. We hope our findings with trigger high-pressure experiments in $Cd_2V_2O_7$.

## Author information

**Corresponding Author**

*daniel.errandonea@uv.es.

**ORCID**

D.Díaz-Anichtchenko: 0000-0002-1263-0291

D. Errandonea: 0000-0003-0189-4221

## Data availability

All relevant data are available from the corresponding author upon reasonable request.

## Author contributions

All authors have contributed equally to this work.

## Conflicts of interest

There are no conflicts to declare.

## Acknowledgments

D. E. thanks the financial support from Generalitat Valenciana under Grant PROMETEO 2018/123-EFIMAT and by the Spanish Research Agency (AEI) and Spanish Ministry of Science and Investigation (MCIN) under projects PID2019-106383GB-C41 (DOI: 10.13039/501100011033) and RED2018-102612-T (MALTA Consolider-Team Network). D. D.-A. acknowledges the PhD fellowship granted by Generalitat Valencia (ACIF/2020/009).

Table captions

**Table 1**. Unit-cell parameters and volume (per formula unit) of different $Cd_2V_2O_7$ structures calculated with B3LYP functional. (a) β-phase, (b) γ-phase and (c) τ-phase. In the case of the β-phase, we also show results calculated with HSE06 and PBE functionals. They are compared with experimental and calculated unit-cell parameters reported in the literature.[17,22]

**Table 2**. Calculated atomic positions for the β-, γ-, and τ-phase. Results are from calculations performed using the B3LYP hybrid functional. The results for the β-, and γ-phase are compared with experiments.[17, 22] which are provided in the right side of the table.

**Table 3**. The unit-cell volume ($Å^3$), bulk modulus (GPa), and bulk modulus pressure derivative at ambient pressure determined using a third-order Birch–Murnaghan EOS.

**Table 4**. Eigenvalues, $\lambda_i$, and eigenvectors, $e_{vi}$, of the isothermal compressibility tensor of the β-phase (top), γ-phase (center), and τ-phase (bottom) at 0 GPa.

**Table 5**. Average of the bond-length, coordination number (CN), and distortion index of different polyhedra for each phase. Values obtained using the program VESTA.[40]

| (a) | B3LYP | HSE06 | PBE | Exp[17] |
|---|---|---|---|---|
| a (Å) | 7.0883 | 7.0504 | 7.0901 | 7.088 |
| b (Å) | 8.4638 | 8.3298 | 8.2798 | 9.091 |
| c (Å) | 4.9885 | 4.9543 | 5.0197 | 4.963 |
| β (º) | 100.2022 | 99.8203 | 99.4645 | 103.350 |
| $V_0$/pfu(Å$^3$) | 147.2748 | 143.3477 | 145.3350 | 155.579 |

| (b) | B3LYP | Exp[22] |
|---|---|---|
| a (Å) | 6.5381 | 6.5974 (2) |
| b (Å) | 6.8351 | 6.8994 (2) |
| c (Å) | 6.9078 | 6.9961 (2) |
| α (º) | 83.775 | 83.325 (1) |
| β (º) | 64.275 | 63.898 (1) |
| γ (º) | 81.141 | 80.145(1) |
| $V_0$/pfu(Å$^3$) | 137.24 | 140.72 (1) |

| (c) | B3LYP |
|---|---|
| a (Å) | 9.9108 |
| $V_0$/pfu(Å$^3$) | 121.6858 |

**Table 1**

### β–phase

| Atom | Site | x_theo | y_theo | z_theo | x | y | z |
|------|------|--------|--------|--------|------|--------|--------|
| $Cd_1$ | 4h | 0 | 0.2912 | 0 | 0 | 0.3052 | 0 |
| $V_1$ | 4i | 0.2344 | 0 | 0.4271 | 0.2262 | 0 | 0.4088 |
| $O_1$ | 2a | 0 | 0 | 0.5 | 0 | 0 | 0.5 |
| $O_2$ | 4i | 0.6091 | 0 | 0.2793 | 0.6119 | 0 | 0.2871 |
| $O_3$ | 8j | 0.2523 | 0.1649 | 0.2423 | 0.2378 | 0.1566 | 0.2199 |

### γ-phase

| Atom | Site | x_theo | y_theo | z_theo | x | y | z |
|------|------|--------|--------|--------|--------|--------|--------|
| $Cd_1$ | 2i | 0.2388 | 0.3394 | 0.8358 | 0.2421 | 0.3367 | 0.8326 |
| $Cd_2$ | 2i | 0.7565 | 0.0450 | 0.7590 | 0.7498 | 0.0344 | 0.7575 |
| $V_1$ | 2i | 0.7035 | 0.1638 | 0.2632 | 0.7104 | 0.1645 | 0.2586 |
| $V_2$ | 2i | 0.2291 | 0.4495 | 0.3434 | 0.2284 | 0.4552 | 0.3441 |
| $O_1$ | 2i | 0.8570 | 0.3277 | 0.0791 | 0.8612 | 0.3328 | 0.0816 |
| $O_2$ | 2i | 0.8545 | 0.0411 | 0.3945 | 0.8622 | 0.0439 | 0.3907 |
| $O_3$ | 2i | 0.4554 | 0.2914 | 0.4410 | 0.4592 | 0.2893 | 0.4363 |
| $O_4$ | 2i | 0.6465 | 0.9937 | 0.1300 | 0.6546 | 0.9936 | 0.1233 |
| $O_5$ | 2i | 0.2625 | 0.2863 | 0.1699 | 0.2714 | 0.2948 | 0.1660 |
| $O_6$ | 2i | 0.3901 | 0.6353 | 0.2328 | 0.3839 | 0.6438 | 0.2436 |
| $O_7$ | 2i | 0.9561 | 0.5925 | 0.3681 | 0.9504 | 0.5892 | 0.3678 |

### τ-phase

| Atom | Site | x_theo | y_theo | z_theo |
|------|------|--------|--------|--------|
| $Cd_1$ | 16d | 0.625 | 0.625 | 0.625 |
| $V_1$ | 16c | 0.125 | 0.125 | 0.125 |
| $O_1$ | 48f | 0.0624 | 0.25 | 0.25 |
| $O_3$ | 8b | 0.5 | 0 | 0 |

**Table 2**

| Sample | Cd-O | Distortion | CN | V-O | Distortion | CN |
|---|---|---|---|---|---|---|
| β | 2.2911 | 0.0675 | 6 | 1.7039 | 0.01724 | 4 |
| γ$_1$ | 2.3400 | 0.02257 | 7 | 1.7053 | 0.01294 | 4 |
| γ$_2$ | 2.3210 | 0.02216 | 6 | 1.8044 | 0.06319 | 5 |
| τ | 2.4514 | 0.06234 | 8 | 1.8586 | 0 | 6 |

Table 3

| Phase | $V_0(\text{Å}^3)$ | $B_0(GPa)$ | $B_0'$ |
|---|---|---|---|
| β | 147.31 | 81.2 | 3.8 |
| γ | 274.39 | 125.2 | 3.9 |
| τ | 243.50 | 226.1 | 4.3 |

Table 4

| | |
|---|---|
| $\lambda_1 = 9.80(5)\ 10^{-3}$ GPa$^{-1}$ | $e_{v1} = (0,-1,0)$ |
| $\lambda_2 = 2.97(2)\ 10^{-3}$ GPa$^{-1}$ | $e_{v2} = (-1,0,2)$ |
| $\lambda_3 = -1.68(4)\ 10^{-3}$ GPa$^{-1}$ | $e_{v3} = (1,0,1)$ |
| $\lambda_1 = 4.16(2)\ 10^{-3}$ GPa$^{-1}$ | $e_{v1} = (3,4,2)$ |
| $\lambda_2 = 2.95(2)\ 10^{-3}$ GPa$^{-1}$ | $e_{v2} = (2,1,-3)$ |
| $\lambda_3 = 0.36(1)\ 10^{-3}$ GPa$^{-1}$ | $e_{v3} = (14,-12,-1)$ |
| $\lambda = 1.33(1)\ 10^{-3}$ GPa$^{-1}$ | |

Table 5

Figure captions

**Figure 1:** Difference of enthalpy of different crystal structures with respect to the to β-phase calculated as a function of pressure using the B3LYP functional. In the bottom horizontal axis, we show the calculated pressure ($P_{calc}$). In the top horizontal axis, we show the pressure (P) obtained after applying the 4 GPa offset discussed in the text.

**Figure 2:** Perspective and projections of the crystal structure of β-phase and representation of the $VO_7$ dimer.

**Figure 3:** Perspective and projections of the crystal structure of γ-phase and representation of the $V_4O_{14}$ anion.

**Figure 4:** Perspective and projections of the crystal structure of τ-phase.

**Figure 5:** Unit-Cell parameters of the β-, γ- and τ-phase versus the pressure. The results shown have been calculated using the B3LYP hybrid functional. In the bottom horizontal axis, we show the calculated pressure ($P_{calc}$). In the top horizontal axis, we show the pressure (P) obtained after applying the 4 GPa offset discussed in the text.

**Figure 6:** Pressure dependence of the unit-cell volume of the β-, γ- and τ-phases of $Cd_2V_2O_7$. The results shown have been calculated using the B3LYP hybrid functional. In the bottom horizontal axis, we show the calculated pressure ($P_{calc}$). In the top horizontal axis, we show the pressure (P) obtained after applying the 4 GPa offset discussed in the text.

**Figure 7:** Band structure and DOS calculated at ambient pressure using the B3LYP hybrid functional. (a) β-phase, (b) γ-phase, and (c) τ-phase.

**Figure 8:** Evolution of the band-gap energy with pressure for the β-, γ- and τ-phases. The results shown have been obtained employing the B3LYP hybrid functional. In the bottom horizontal axis, we show the calculated pressure ($P_{calc}$). In the top horizontal axis, we show the pressure (P) obtained after applying the 4 GPa offset discussed in the text.

Figures

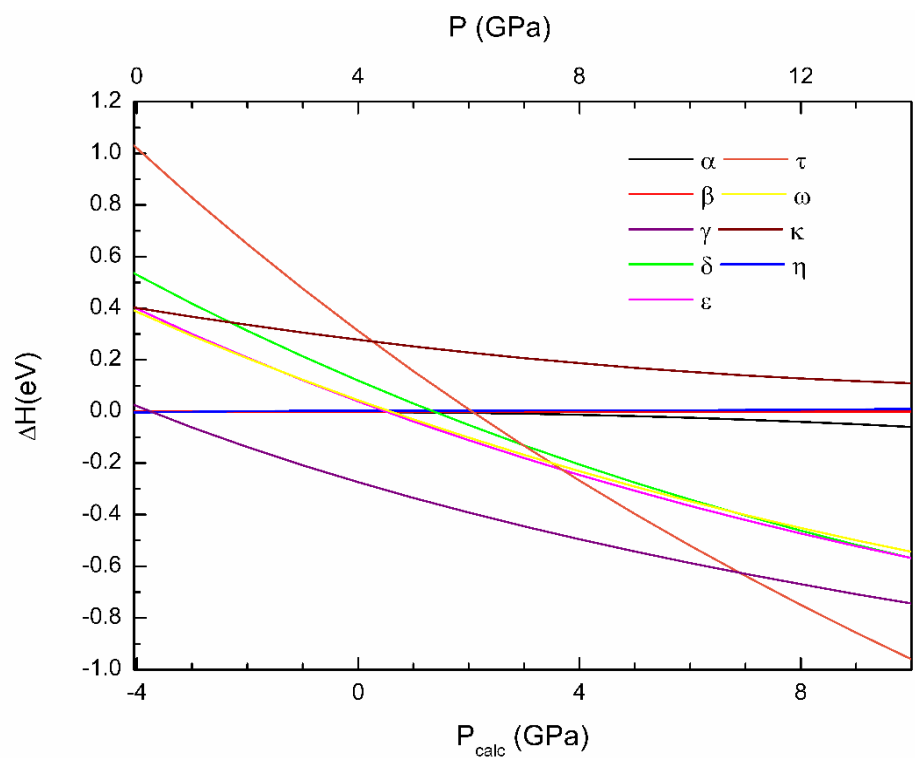

**Figure 1**

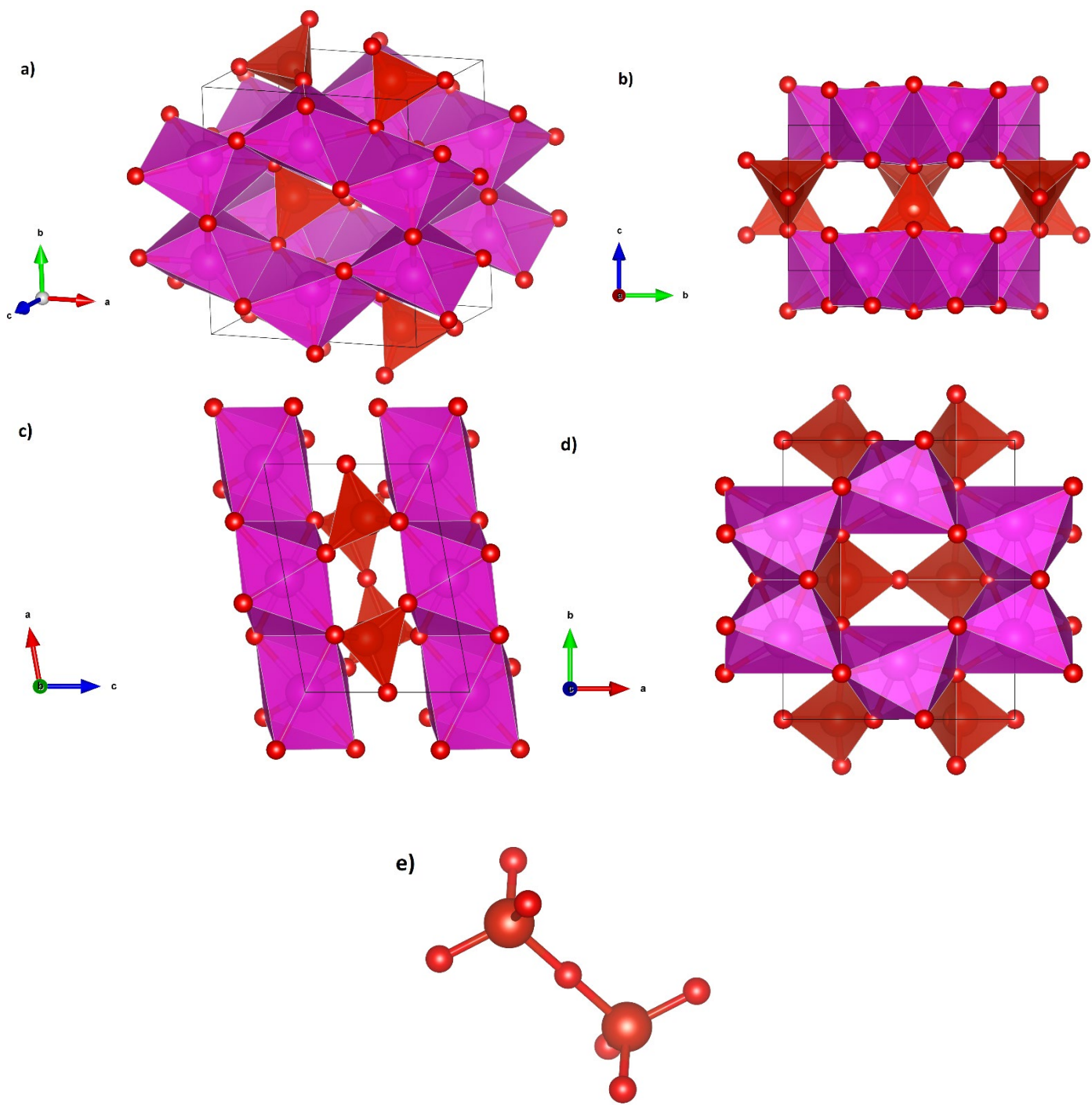

**Figure 2**

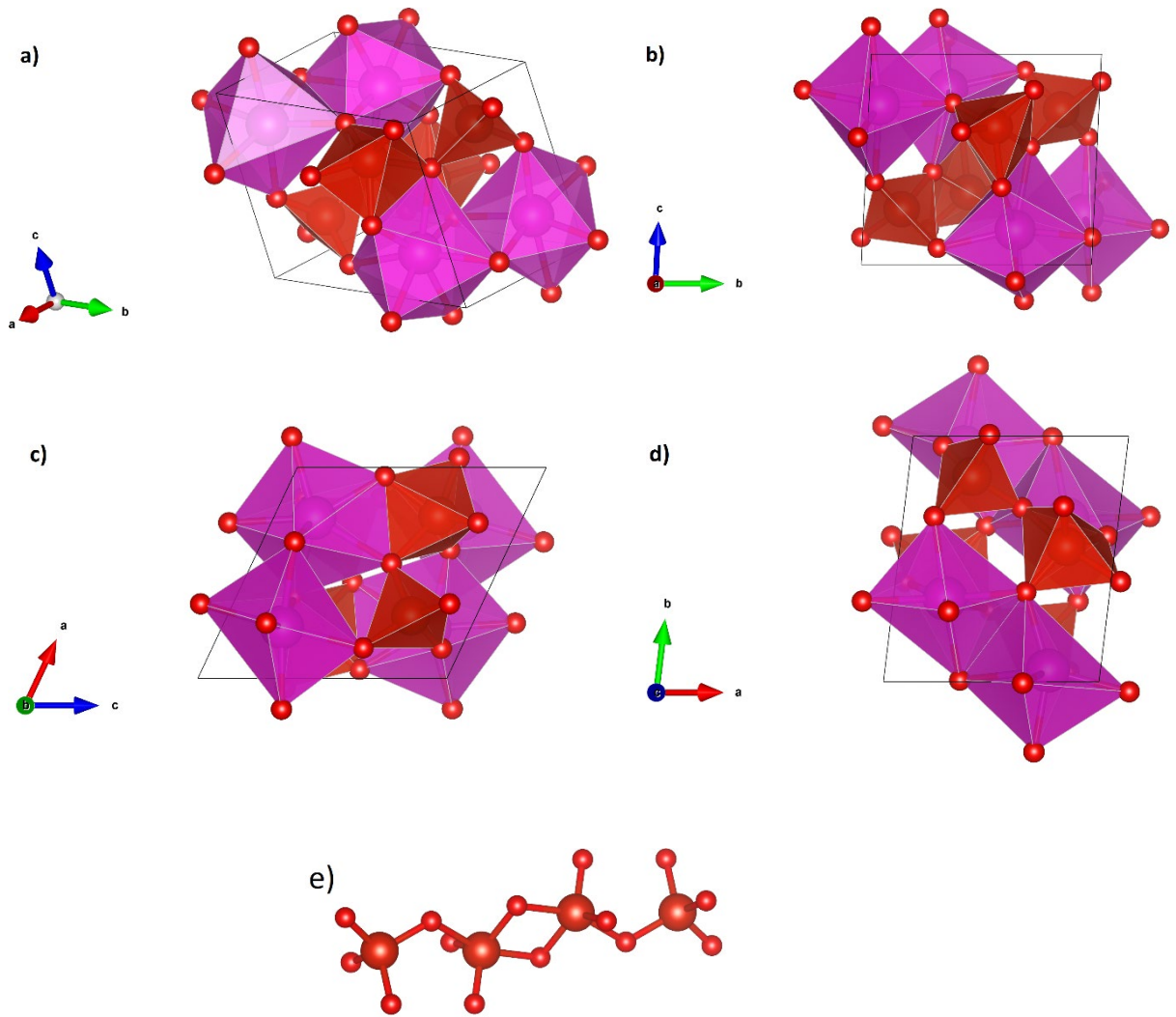

**Figure 3**

a)

b)

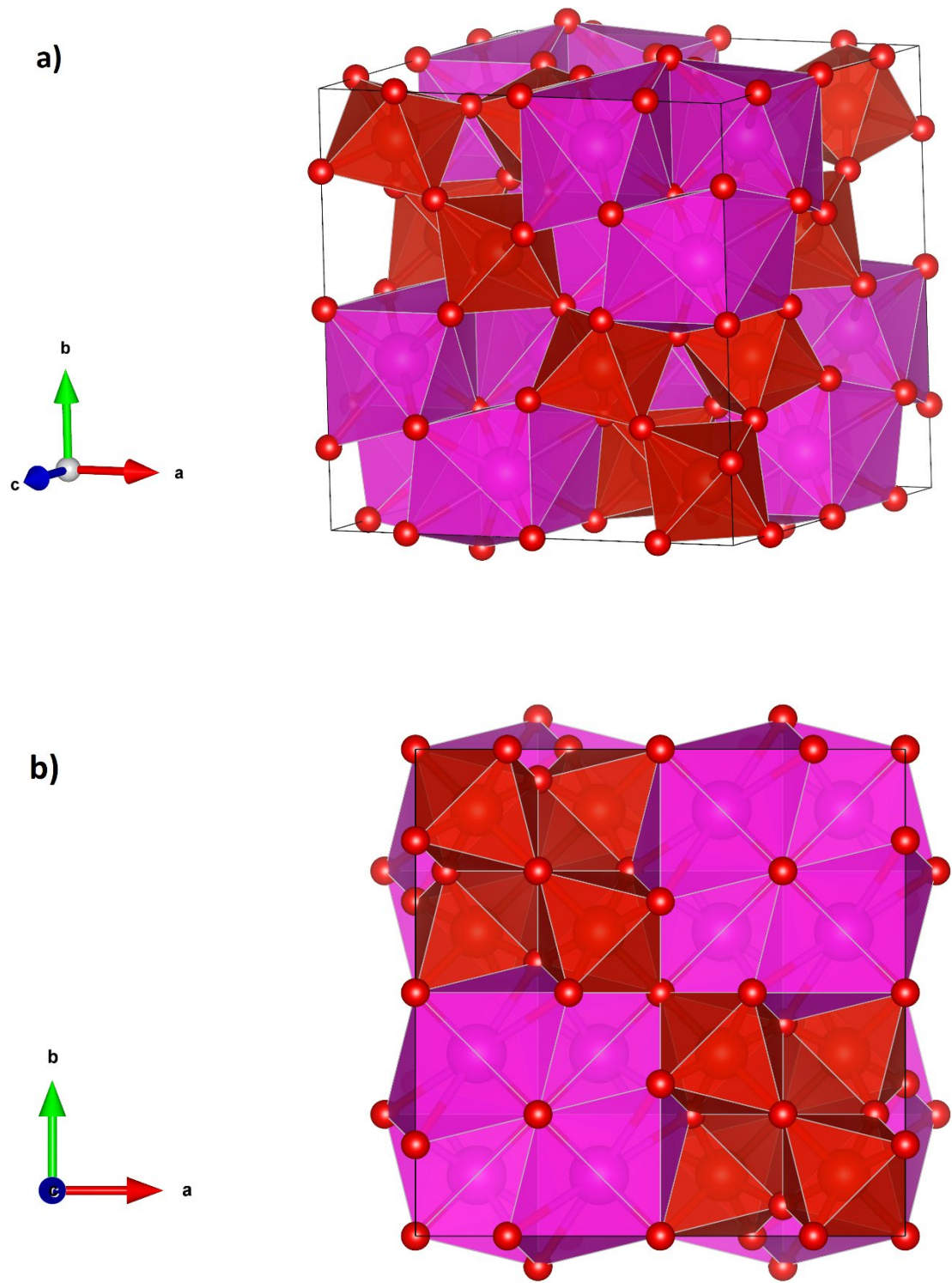

**Figure 4**

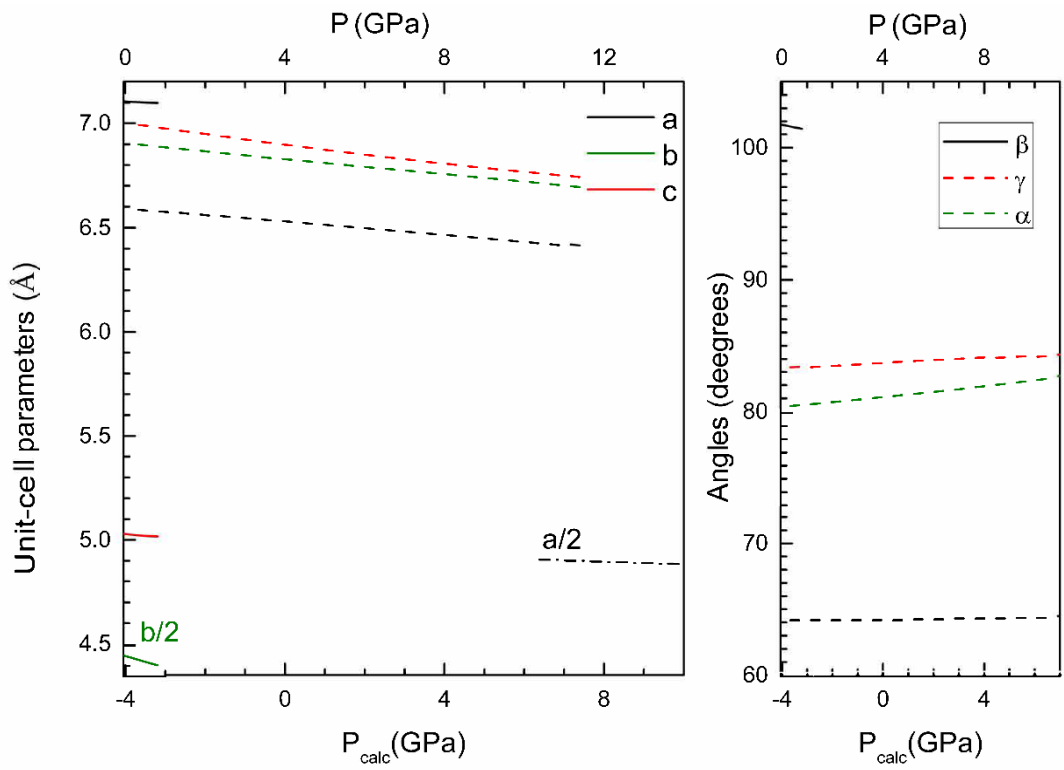

**Figure 5**

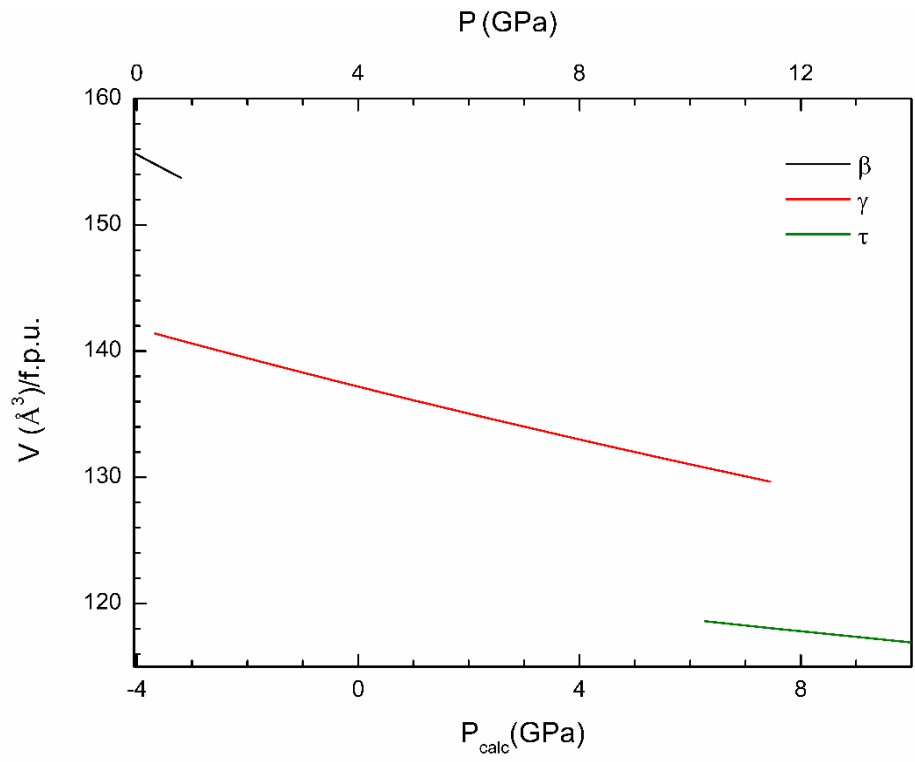

**Figure 6**

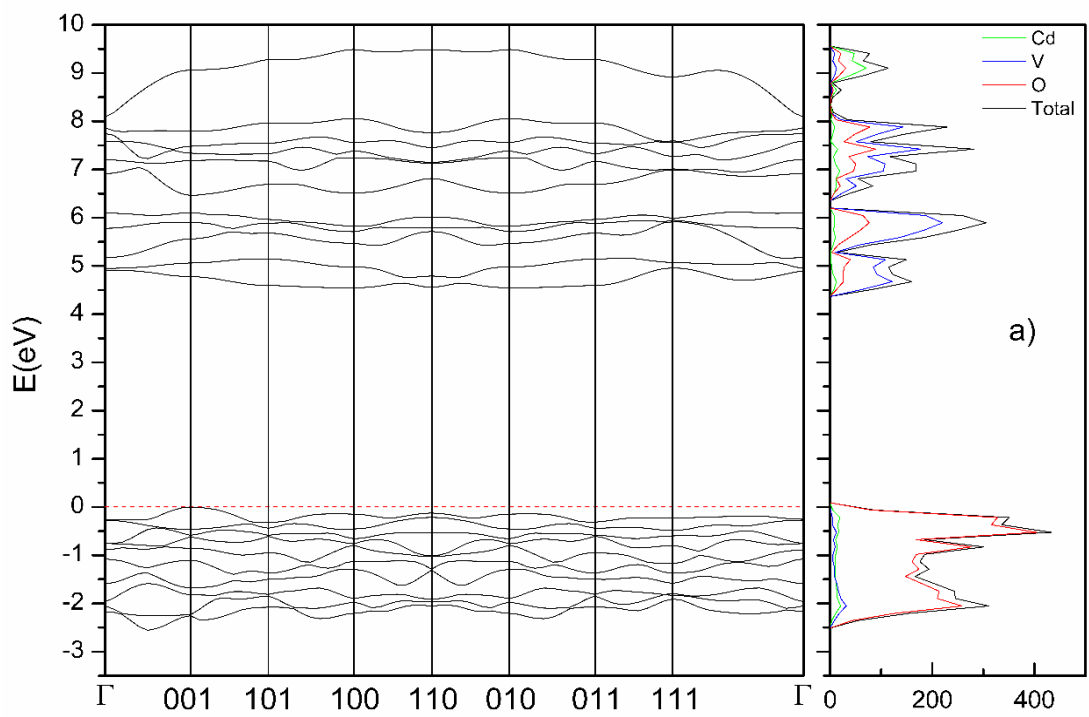

**Figure 7(a)**

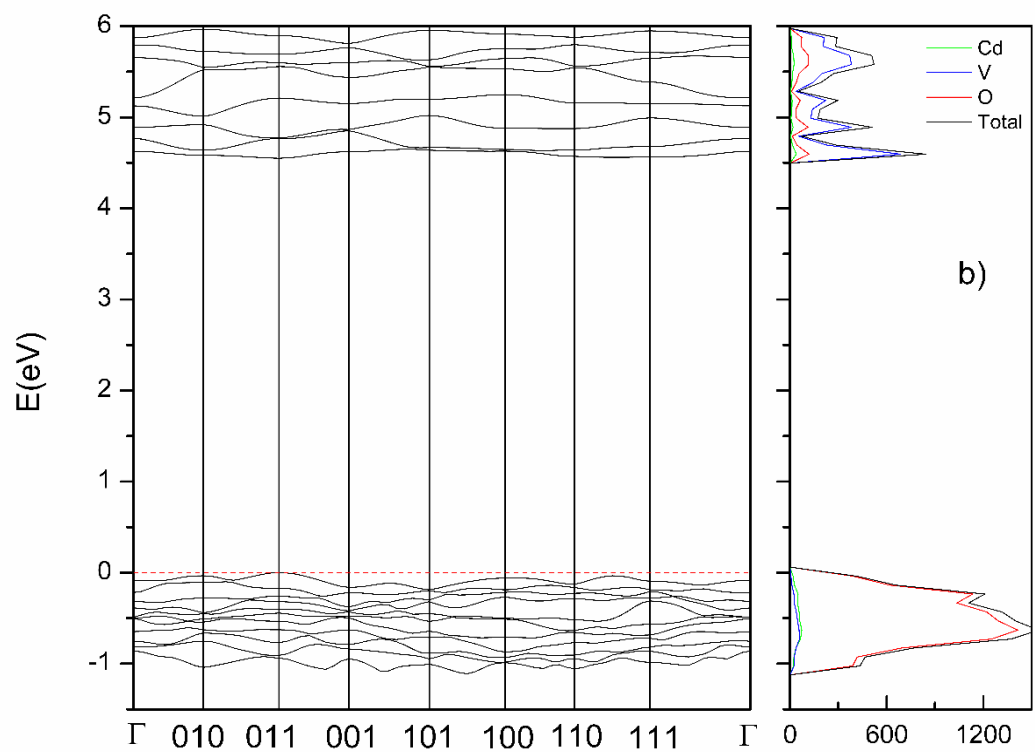

**Figure 7(b)**

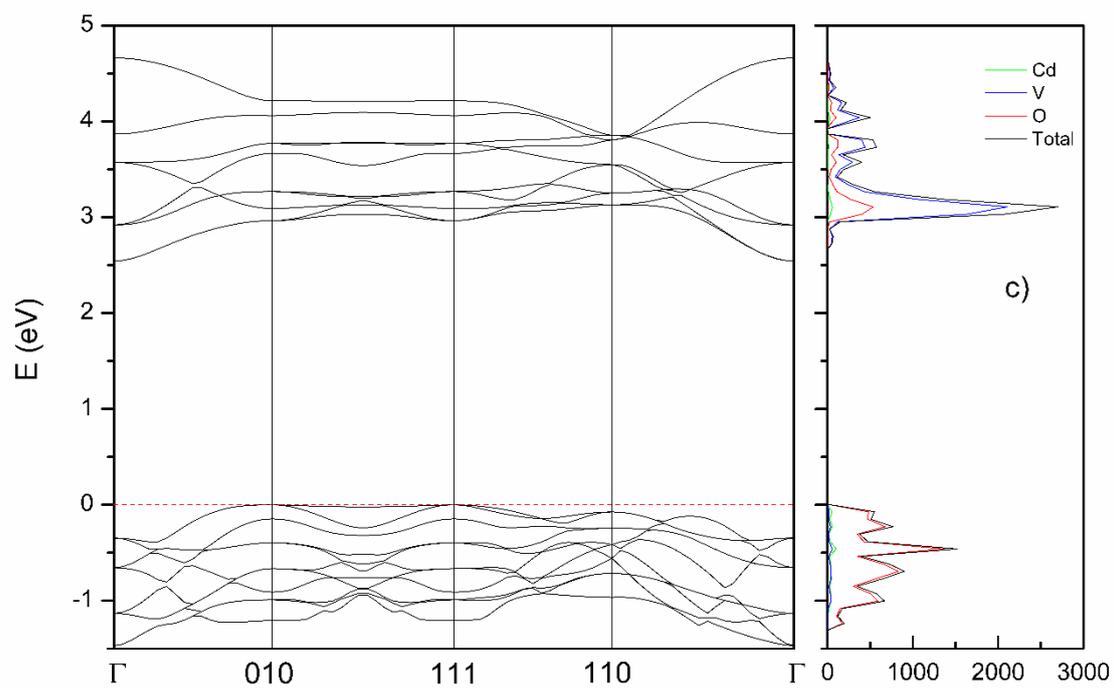

**Figure 7(c)**

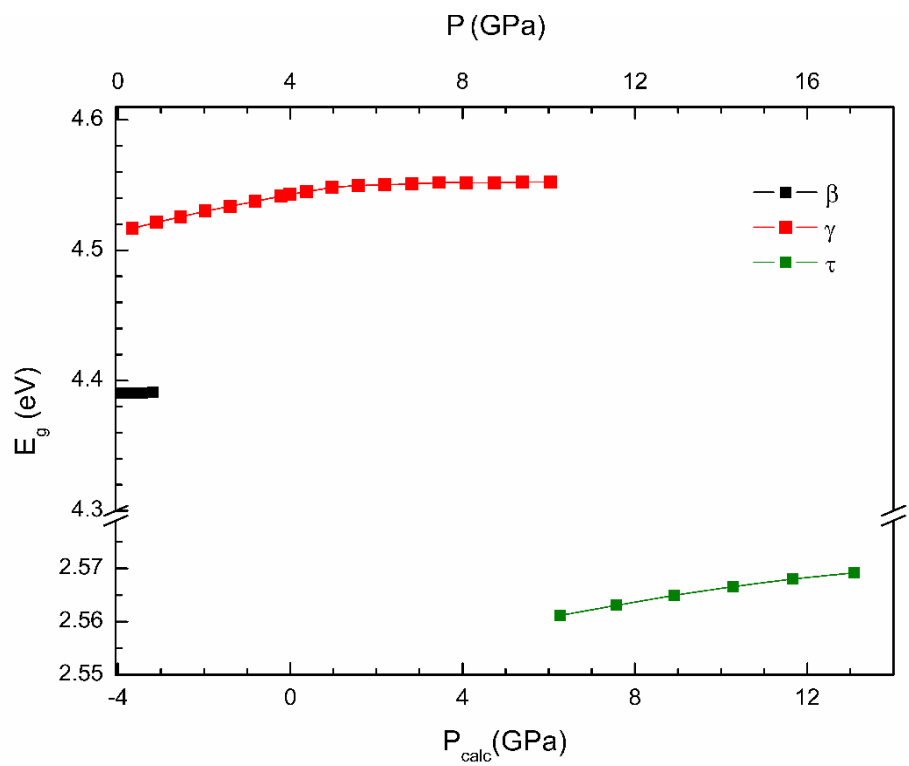

**Figure 8**